# Multiscale distribution of oxygen puddles in 1/8 doped $YBa_2Cu_3O_{6.67}$

Alessandro Ricci[1,2], Nicola Poccia[2,3], Gaetano Campi[4], Francesco Coneri[3], Alessandra Stella Caporale[2], Davide Innocenti[2], Manfred Burghammer[5], Martin v. Zimmermann[1], Antonio Bianconi[2,*]

[1]*Deutsches Elektronen-Synchrotron DESY, Notkestraße 85, D-22607 Hamburg, Germany.*

[2]*Rome International Center for Materials Science Superstripes RICMASS, via dei Sabelli 119A, 00185 Roma, Italy*

[3]*MESA+ Institute for Nanotechnology, University of Twente, P. O. Box 217, 7500AE Enschede, Netherlands*

[4]*Institute of Crystallography, CNR, via Salaria Km 29.300, Monterotondo Stazione, Roma, I-00016, Italy*

[5]*European Synchrotron Radiation Facility, B. P. 220, F-38043 Grenoble Cedex, France*

*Correspondence to: antonio.bianconi@ricmass.eu

Despite intensive research a physical explanation of high $T_c$ superconductors remains elusive. One reason for this is that these materials have generally a very complex structure making useless theoretical models for a homogeneous system. Little is known on the control of the critical temperature by the space disposition of defects because of lack of suitable experimental probes. X-ray diffraction and neutron scattering experiments used to investigate $y$ oxygen dopants in $YBa_2Cu_3O_{6+y}$ lack of spatial resolution. Here we report the spatial imaging of dopants distribution in-homogeneity in $YBa_2Cu_3O_{6.67}$ using scanning nano X-ray diffraction. By changing the X-ray beam size from 1 micron to 300 nm of diameter, the lattice inhomogeneity increases. The ordered oxygen puddles size distribution vary between 6-8 nm using 1x1 $\mu m^2$ beam, while it is between 5-12 nm with a fat tail using the 300x300 $nm^2$ beam. The increased inhomogeneity at the nanoscale points toward an intrinsic granular complexity.

Physical Sciences: Applied Physics

While the layered heterostructure at atomic limit is accepted as a universal feature for the lattice of high temperature superconductors, little is known on the control of the critical temperature by the space disposition of defects in the spacer layers[1-7]. Defects, as mobile oxygen interstitials, can control electronic inhomogeneity[8-14] and complex multi-scale phase separation[15-24]. Furthermore, inhomogeneity in the form of stripes have been shown to induce superconducting shape resonances and amplification of gaps and critical temperature driving the BCS-BEC crossover in the system, thanks to a strong renormalization of the chemical potential.

The advances in control of atomic defects and their spatial self-organization is needed for the engineering of new electronic devices at atomic level[11]. Imaging is needed to control defects organization at the nanoscale in novel quantum functional superconducting materials and it is the first step for unveiling the structure–property relation. Several spatial length scales are involved, which creates strongly intermingled defects structures. In high temperature superconductors (HTS), these networks of defects can influence essential properties such as the Fermi surface, pseudogap energy, spin density waves, charge density waves, and even the superconducting critical temperature. The strong tendency towards complex phase separation[14-24] can be used as advantage for developing new electronic devices. In fact, in some circumstances, defects order can be manipulated by scanning tunneling microscopy (STM)[11], and/or by continuously exposing the sample to a UV or X-ray beam[6,7,10]. Developing techniques for imaging the nanoscale heterogeneity of defects organization is the first step to open new opportunities for their manipulation.

$YBa_2Cu_3O_{6+y}$ (YBCO) high temperature superconductor is one of the most studied HTS material because of its superconducting transition above the nitrogen liquid evaporation temperature. From high resolution hard X-ray diffraction and neutron scattering experiments it is known that the *y* oxygen dopants form one-dimensional oxygen-compositional stripes that segregate in short-range puddles[25-30] but few information exist on their spatial distribution. Here we report a spatial imaging using scanning nano X-ray diffraction [5-9] of the spatial distribution of the puddles made of compositional oxygens stripes (Ortho-VIII puddles) in the basal plane of $YBa_2Cu_3O_{6.67}$ at hole doping $p \approx 1/8$[31-34].

In particular high quality untwined single crystals of YBCO with doping set at 6.67 corresponding to 1/8 holes per Cu site in the $Y(CuO_2)_2$ bilayer, have become available in these last years[26]. These new crystals have allowed the identification of the Fermi surface reconstruction with the appearance of a new electron-pocket[32], the anomalous Nerst effect[32], the proximity to a critical hole doping regime[33], the unusual magnetic orders[34] and broken symmetries[35]. Although the observation of all these effects is certainly due to the high quality of the samples, understanding the disposition in real space of the YBCO structural components could be a difficult task, as we may expect form the variety of its oxygen phases in the phase diagram[26-29]. Of course, as the oxygen ions and the charged vacancies order, a corresponding charge density wave (CDW) is generated as well.

The key result of this experiment is to unveil the intrinsic inhomogeneity of oxygen puddles distribution. We have recorded images with different spatial resolution by changing the X-ray beam size from 1 micron to 300 nm of diameter. The Ortho-VIII puddles size normal distribution is in the range 6-8 nm for the 1x1 µm$^2$ beam while it is much broader (in the range 5-12 nm) with a fat tail using the 300x300 nm$^2$ beam. The increased inhomogeneity at the nanoscale points toward an intrinsic granular complexity of puddles at 1/8 doping. This is confirmed by the correlation lengths of the spatial correlation function of the maps that show a dramatic reduction decreasing the beam size.

## Results

Synchrotron radiation x-ray diffraction shows the high quality of the YBa$_2$Cu$_3$O$_{6.67}$ single crystal with P4/m spatial symmetry, unit cell dimensions a=3.807(11) Å, b=3.864(12) Å, c=11.52(2) Å and volume of 169.5(8) Å$^3$. The diffuse streaks of the superlattice reflections have no well-defined peaks along *l* and the streaks shown in **Fig. 1A and B** provide direct evidence for nanoscale puddles in the basal *a-b* plane with Ortho-VIII modulation with a substantial disorder in the stacking of full and empty chains along the *c* direction. The nanoscale size of Ortho-VIII puddles embedded in a disordered medium is shown as pictorial view in **Fig. 1C** representing the basal Cu-O plane of YBCO. The Ortho-VIII puddles are made of alternated 5 filled and 3 empty oxygen wires every 8 rows. The number of oxygen ions per unit in the puddles is 6.625= 6.5+0.125 that gives exactly 1/8 holes per Cu site in the Y(CuO$_2$)$_2$ bi-layers with Cu$^{2+}$ ions in the chains of the basal plane. Since the average oxygen concentration is 6.67 =6.5+1/6 the hole doping in the disordered background is larger than 1/8.

In order to understand the real space disposition of the Ortho-VIII puddles, and to avoid the transmission electron microscopy (TEM) complications due to electron beam damage of the sample, we have used scanning micro/nano X-ray diffraction (µXRD/nXRD) developed at the ID13 beam-line of the European Synchrotron Radiation Facility in Grenoble (France) (see supplementary info). µXRD and nXRD techniques have been already successfully applied on La214,[5-7] Bi2212[8] cuprates and on iron-based superconductors[9].

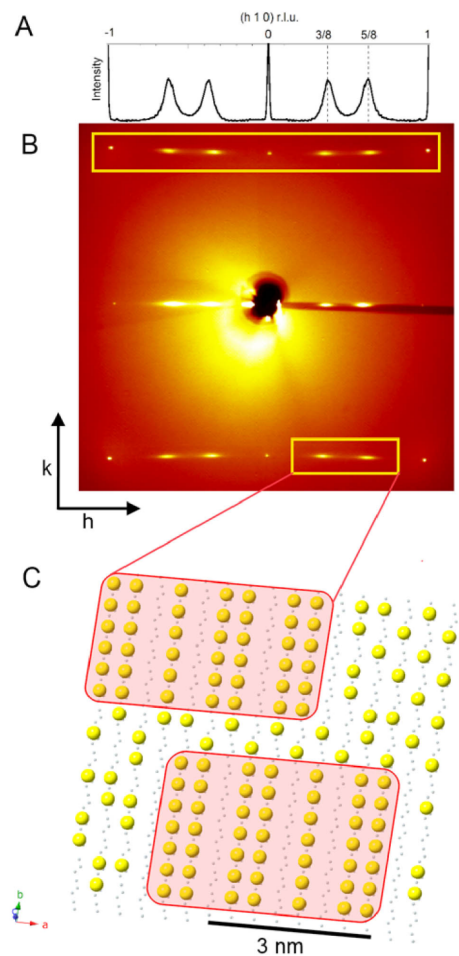

**Figure 1.** $YBa_2Cu_3O_{6.67}$ single crystal diffraction pattern at room temperature. **(a)** The X-ray diffraction pattern shows diffuse superstructure satellites at 3/8 and 5/8 positions, their profile along *h* is plotted. **(b)** Room temperature diffraction pattern collected using a transmission geometry. The CuO chains superlattice reflections $q_{Ortho-VIII}(a^*)=(3/8,k,0)$ and $q_{Ortho-VIII}(a^*)=(5/8,k,0)$ are highlighted in the empty yellow box. **(c)** Pictorial view of the basal Cu-Oiy plane of YBCO. The small grey dots are the Cu(1) sites and the large yellow dots are the oxygen ions Oiy in defective basal plane. The pictorial view of oxygen ordered metamorphic phases called Ortho-VIII, with formal oxygen content $y_{puddle}=0.625$ shown in the pink filled rectangles, are embedded in the disordered background with an oxygen content slightly larger than the average oxygen content $y_{aver}=0.67$.

In this work by scanning micro areas, this technique gives a mixed information of the k- and r-space of the bulk structure inhomogeneities and it has never been applied on a $YBa_2Cu_3O_{6.67}$ single crystal. A 1x1 µm$^2$ and 300x300 nm$^2$ beam size, have been employed for our measurements at room temperature. The integrated intensity of the observed superstructure in **Figure 2A** and the peak position of satellites in different spot of the crystal are quite homogenous, $q_{ortho-VIII}$ = 0.625 with a standard deviation of 0.001. This indicates that the nanoscale oxygen puddles have the Ortho-VIII periodicity and confirms the high quality of the crystal. On the contrary, the FWHM along a* (**Figure 2B**) and along c* (**Figure 2C**) of diffraction profiles, measured at each point (x,y) of the sample reached by the x-y translator with micron resolution, clearly show a spatial variation.

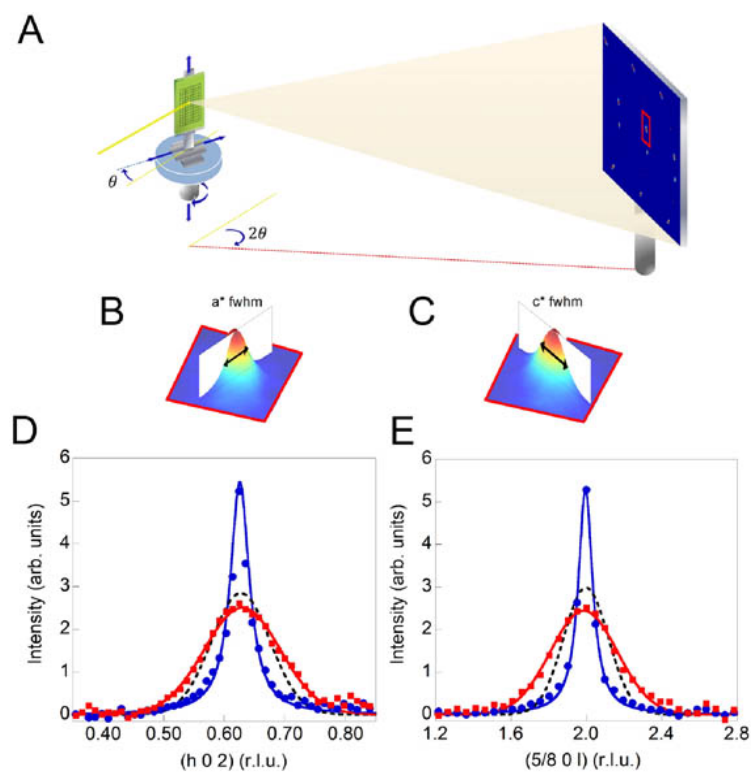

**Figure 2.** (a) Scanning micro X-ray diffraction experimental set-up at the ID13 beamline of the European Synchrotron Radiation Facility (ESRF) of Grenoble. Full width half maximum (FWHM) of the superlattice reflections along h $q_{Ortho-VIII}(a^*)$=(h,0,2) **(b)** and along l $q_{Ortho-VIII}(c^*)$=(5/8,0,l) **(c)**. The (h 0 2) and the (5/8 0 l) superlattice peak profile respectively along a* **(d)** and along c* **(e)**. The blue filled dots, the black dotted lines and the red squares correspond respectively to the profile of a narrow, an average and a broad reflection. Using therefore a micron and sub-micron X-ray beam we are able to resolve the peaks FWHM fluctuations from one spot to the other in the sample.

The diffraction profiles (**Figure 2D, 2E**) show the position dependence of the domain size, derived from the measured full width half maximum (FWHM) of the superstructure reflections via standard methods of diffraction. The spatial variation of the domain size in the *a* and *c* crystallographic directions is showed for three spots of the sample corresponding respectively to a narrow, average and a broad superstructure reflection. Although the micro X-ray diffraction profiles give a rough picture on the size variations of the Ortho-VIII puddles, three spots cannot give enough information to construct a real space image of the organization of the Ortho-VIII puddles.

To solve this issue, we have collected 10000 images of µXRD diffraction patterns, scanning the sample with a 1 micron step along both the real space direction x-y (*a-c* plane). **Figure 3A** and **Figure 3B** show the maps of the Ortho-VIII domain size along a* and along c*. Although the color bar of the maps extends from dark blue to dark red and is associated to 5 nm to 12 nm sizes respectively, actually the observed color are green and yellow, which indicate a smooth variation of the puddles size in the system within a range of 6.5 nm to 8.5 nm.

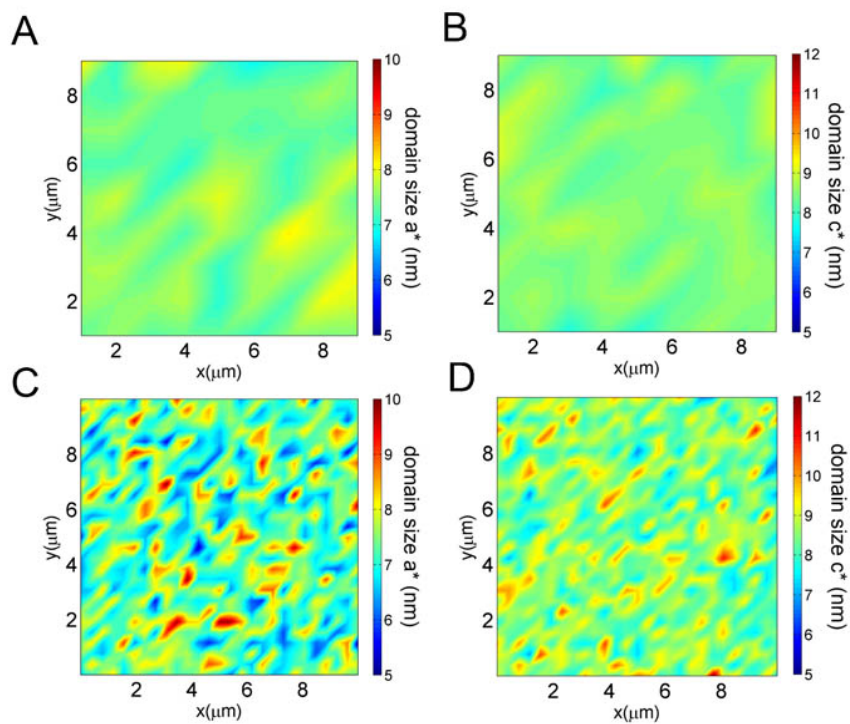

**Figure 3** The position dependence of the Ortho-VIII domain size along a***(a)** and along c* **(b)**, obtained from data measured by scanning micro X-ray diffraction, using a 1 micron incident beam on the sample. The Ortho-VIII domain size do have variations which range from light blue to yellow in a region of about 9 micron squares. The same region size has been investigated, using a 300 nm incidence beam on the sample. The Ortho-VIII mapping of the domain size along a* **(d)** and along c* **(e)** are shown. Here the variation is more pronounced, indeed, the intense red–yellow peaks in the two-dimensional color map represent locations in the sample with high strength of Ortho-VIII ordering, and dark blue indicates spots of disordered Ortho-VIII domains.

The average size of this interval is in agreement with the average size of the Ortho-VIII puddles investigated with hard X-ray diffraction at ELETTRA using a 200x200 µm² beamsize. To test if the observed size variation of the Ortho-VIII puddles is an intrinsic aspect of the YBCO system or it is a sample dependent property, we have increased the spatial resolution of the X-ray beam, by diminishing the beam size and performed the same experiment in a similar area. Using a monochromatic X-ray beam of photon energy 14 KeV ($\Delta E/E = 10^{-4}$), the beam has been focused by Kirkpatrick Baez (KB) mirrors down to a 300 nm spot size on the sample (full width at half maximum). We collected other 20000 diffraction images in the reflection geometry using a Fast Readout Low Noise charged coupled device (FReLoN CCD) detector.

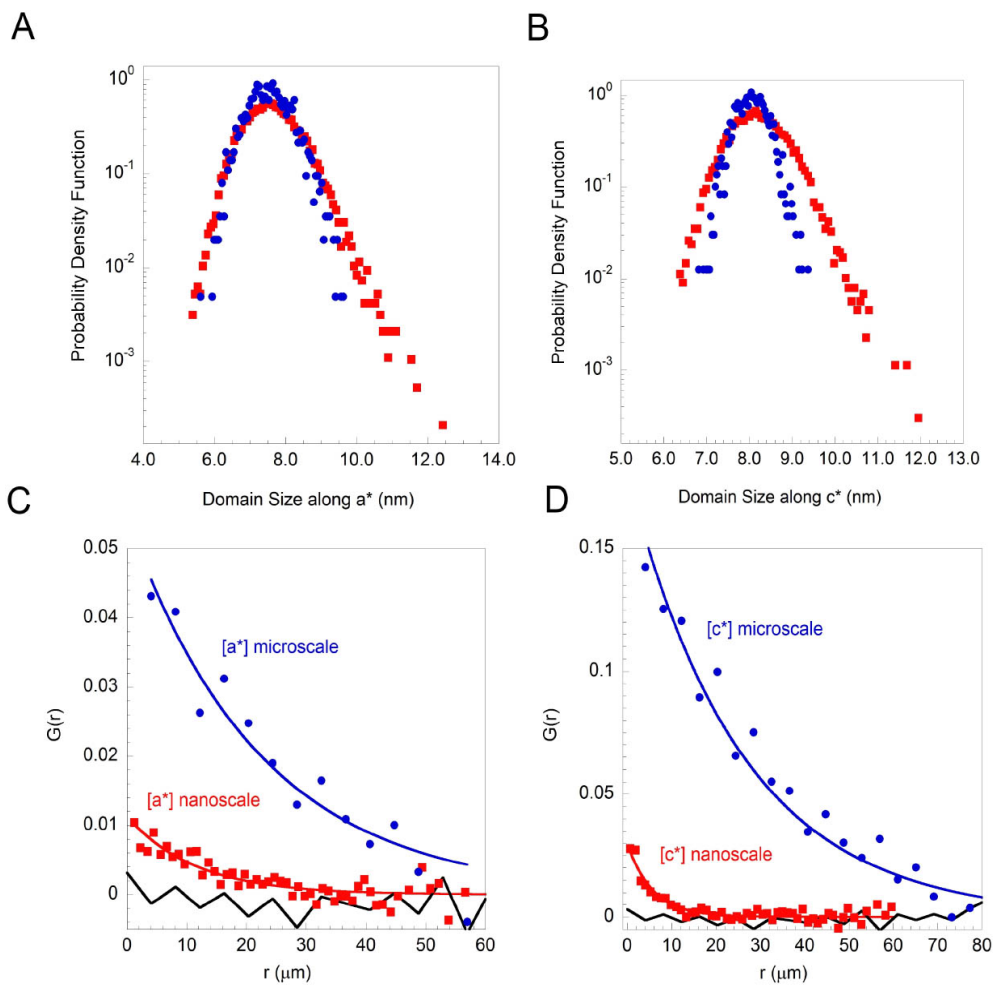

**Figure 4** Probability density function of the domain size along a-axis **(a)** and along c-axis **(b)** obtained by the microbeam (filled circles) and the nanobeam (filled squares) measurements. Radial correlation function G(r) calculated on domain size map along a-axis **(c)** and c-axis **(d)**. The blue circles and the red filled squares correspond to the µXRD and nXRD measurements. A comparison between the G(r) behaviors is shown.

We constructed the 2D maps from the collected nXRD diffraction patterns, each one for a different spatial x-y position (*a-c* plane) of the sample. **Figure 3C** and **3D** show respectively the 2D plots of the domains size variation along *a* and *c* crystallographic directions. Visual inspection of the maps show a more inhomogeneous spatial distribution of the Ortho-VIII puddles respect that measured delivering a 1x1 µm$^2$ beam on the sample (µXRD case). Here the color extends from dark blue to dark red, respectively associated to 5 and 12 nm. The broad distribution of sizes shown in the map indicates an intrinsic and incommensurate nanoscopic order in the system. In addition, the heterogeneous granular structure of the dopants, unveiled by the employ of a nanobeam, shows the relevance of the spatial resolution of the probe on the sample.

Statistical analysis is used to quantify the size distributions of the Ortho-VIII puddles as seen by scanning micro and nano X-ray diffraction experiments. **Figure 4A** shows the probability density function (PDF) of the Ortho-VIII puddles sizes along the crystallographic a-axis, observed using a beam of 1 µm and 300 nm of diameter. The distribution of the puddles is nearly symmetric as seen with a resolution of 1 micron while with 300 nm X-ray beam it shows a remarkable asymmetric fat tail, although the average value of 7.5± 0.2 nm is comparable. **Figure 4B** shows the probability density function of the Ortho-VIII puddles sizes along the c crystallographic direction. Also in the size of the puddles in the transversal direction from the $CuO_2$ planes is nearly symmetric however the distribution of sizes measured by the 300 nm nXRD probe, shows a remarkable fat tail.

Also in this case the maps obtained by µXRD and the nXRD are in agreement for the average size of domains which is 8.0 ± 0.2 nm. The standard deviation is 0.4 nm for the 1 micron X-ray probe and 0.6 nm for the 300 nm X-ray probe. The fat tails, observed in the maps, have been quantified by the evaluation of the higher statistical moments (**see supplementary info**). The third moment called Skewness $\gamma$ gave $\gamma_a$= 0.47 and $\gamma_c$= 0.34 for the Ortho-VIII domain size along the a and c direction, respectively. Their positive values indicate a larger weight of the tails in both distributions; in the maps measured with the 1 micron X-ray probe, the Skewness values $\gamma_a$ = 0.23 and $\gamma_c$ = 0.09 result quite smaller (**see Tab. 1**). Similar situation is evidenced evaluating the fourth statistical moment, the Kurtosis $\kappa$, that is an indication of the distribution sharpness respect to the Gaussian shape that has $\kappa$=0. Like is summarized in **Tab. 1** $\kappa_a$= 0.61 and $\kappa_c$= 0.64 for the Ortho-VIII domain size measured by nXRD and $\kappa_a$= 0.03 and $\kappa_c$= 0.05 for µXRD results

| PDF(x) Moments | Nano-beam 300x300 nm$^2$ | | Micro-beam 1x1 µm$^2$ | | Gaussian case |
|---|---|---|---|---|---|
| | *Size-a* | *Size-c* | *Size-a* | *Size-c* | |
| *Skewness (γ)* | 0.47(1) | 0.34(1) | 0.23(1) | 0.09(1) | 0 |
| *Excess of Kurtosis (κ)* | 0.61 (1) | 0.64(1) | 0.03(1) | 0.05(1) | 0 |
| *G(r) exponent (ξ) in µm* | 11.0(2) | 5.1(2) | 22.0(2) | 25.0(2) | |

**Table 1** Statistical moments of the Ortho-VIII domain size PDF along *a* and *c* direction. PDFs are calculated from the intensity maps measured by µXRD (beamsize of 1x1 µm$^2$) and nXRD (beamsize of 300x300 nm$^2$). Both PDFs from µXRD and nXRD show compatible mean values, but variance increased a lot going from the first to second case. The third and the fourth statistical moments deviate from the nominal 0 value for a Gaussian distribution. In particular it is evident that the Skewness and Kurtosis parameters increase a lot decreasing the dimension of the incident beam. In addition, the coherence length ξ, calculated from G(r), is shown in the last row, going from micro-beam to nano-beam it shows a dramatic reduction indicating the intrinsic in-homogeneity of cuprates.

In all the cases, the Skewness and the Kurtosis parameters, result to deviate from the Gaussian distribution characteristic 0 values. The more symmetric distribution observed by the µXRD can be easily explained by the fact that the X-ray beam probes a sample volume of the order of $\mu m^3$, since the effective X-ray penetration depth is in the micron range, that contains about 10$^6$ puddles, on the contrary using the nXRD the number of Ortho-VIII puddles decreases by a factor higher than 10. Although the number of 10$^5$ puddles, in each illuminated spot of the nXRD, is still quite large, it allows to unveil the fat tails of the Ortho-VIII puddles quite clearly.

Looking at the distance-dependent intensity correlations in **Fig 4C** and **4D** measured with a 1 micron and 300 nm X-ray beam, it is clear that along both the directions a* and c* the size of the beam unveils different level of spatial organization. Although the G(r) measured by µXRD and nXRD show that the spatial correlations are more extended out-of-plane than in-plane, the G(r) appears different by comparing the result from different beam sizes. For the domain sizes distribution along the a directions, the correlation length of 22 micron measured by µXRD, is reduced of about the half (to 11 micron) using a 300 nm X-ray beam (nXRD). Along the c directions, the reduction is even more dramatic, in fact, from about 25 micron measured with the 1 micron X-ray beam, the correlation length of the scanned area with a 300 nm X-ray beam is 5 micron.

## Discussion

We have shown that the intrinsic granular organization of mobile oxygen defects can be unveiled by scaling the size of the X-ray beam in the specific case of $YBa_2Cu_3O_{6.67}$ single crystal. If we were seeing a narrowing in the size distributions and a broadening in the correlation length of the puddles, this would have mean that the inhomogeneity observed with the 1 micron beam was due to the sample preparation procedure. Instead, what we measure is a broader and asymmetric distribution of the puddles sizes which show a range of 5-12 nm and a shorter correlation length both along a and c-direction, by scaling the X-ray beam from 1 micron to 300 nm. So far, the defects self-organization and the related nanoscale phase separation controlled by thermal treatments have been well established in high temperature superconductors, like super-oxygenated $La_2CuO_{4+y}$ [5-7], $Bi_2Sr_2CaCuO_{8+y}$ [8] and electron doped iron-chalcogenides[9]. Therefore, the present results support the idea that each high temperature superconductor displays different and specific realizations of a granular and spatially inhomogeneous lattice, controlled by misfit strain[36-38]. Moreover, high temperature superconductivity resists to the complex lattice broken symmetry that is the characteristic feature of the superstripes scenario[39,40]. These results show that also broken symmetry observed in the pseudogap phase[35] could be triggered by a lattice broken symmetry in nanoscale grains and in the domain walls at the interface between the nano-puddles.

We have investigated the spatial arrangement of Ortho-VIII puddles in $YBa_2Cu_3O_{6.67}$, using scanning micro and nano X-ray diffraction for measuring superstructure reflections with unprecedented real space resolution up to 300 nm. Nevertheless the crystal is clearly of high quality, intrinsic inhomogeneities are found in the Ortho-VIII puddles size distributions. In particular, comparing maps obtained from measurements at different spatial scale (1x1 µm$^2$ vs. 300x300 nm$^2$), we are able to show how the inhomogeneities appears more clearly at smaller scale, indicating such local structural inhomogeneities as an intrinsic property in these materials.

# Methods

The high-quality untwined YBCO crystals were grown in a nonreactive $BaZrO_3$ crucible from high-purity starting materials[26]. The oxygen content was set at y=0.67. The hole concentration (doping) p=0.12 was determined from the c-axis lattice constant. The diamagnetic response across the superconductive transition of the $YBa_2Cu_3O_{6.67}$ single crystal has been characterized by means of the Vibrating Sample Magnetometer (VSM) option in a Physical Properties Measurement System (PPMS 6000) from Quantum Design (see supplementary info). After cooling down the sample from room temperature to 10 K in zero applied field (Zero Field Cooling, ZFC) the diamagnetic response upon application of H = 20 Oe has been observed during the warm up across the transition, spanning the 10K to 80K temperature range at a rate of 0.5K/min K. The superconducting transition temperature results to be $T_c$ = 66 K. To study the superlattice reflections due to the oxygen ions (Oi) ordering in the basal plane Ortho-VIII superstructure we performed x-ray diffraction measurements in transmission geometry at the XRD1 beamline of ELETTRA in Trieste (Italy) (see supplementary info) with an energy of 20 KeV and a 200x200 µm² beamsize. We recorded the x-rays scattered by the sample in the a*b* diffraction plane using a charged coupled device (CCD) and results are in agreement with the hard X-ray diffraction data[26]. Scanning micro/nano X-ray diffraction patterns were collected in the reflection geometry with a CCD that records the X-rays scattered from the sample in the *a*c** diffraction plane, showing the Ortho-VIII oxygen chains superlattices $q_{Ortho-VIII}(a^*)=(3/8,k,0)$ and $q_{Ortho-VIII}(a^*)=(5/8,k,0)$.

**Acknowledgments.** We thank Ruixing Liang, D. A. Bonn, and Walter N. Hardy of the Department of Physics of the University of British Columbia for providing us with the crystals. We thank Gabriel Aeppli for helpful discussions and the ID13 beamline staff of ESRF and the XRD1 beamline staff of ELETTRA, especially G. Arrighetti, L. Barba, G. Bais and M. Polentarutti. We are grateful to superstripes onlus for financial support.

**Authors contributions.** A.R., N.P., G.C., M.B. performed the experiment. A.R. G.C., N.P., D.I., A.S.C., A.B. performed the analysis of the data. N.P., F.C. performed the vibrating sample magnetometer measurements. A.B. A.R., N.P., G.C., have conceived and planned the experiment. A.R., A.B., N.P., G.C., M.v.Z. have written the paper.

**Competing financial interests**

The authors declare no competing financial interests.